\documentstyle[12pt,aasms4,overcite,flushrt]{article}
\renewcommand{\thebibliography}[1]{\clearpage\subsection*{REFERENCES}\list
 {\arabic{enumi}.}{\settowidth\labelwidth{[#1]}\leftmargin\labelwidth
 \advance\leftmargin\labelsep
 \usecounter{enumi}}
 \def\newblock{\hskip .11em plus .33em minus .07em}
 \sloppy\clubpenalty4000\widowpenalty4000
 \sfcode`\.=1000\relax}

\def\aa#1{{\it Astr. Astrophys.\/} {\bf #1}}

\def\apj#1{{\it Astrophys. J.\/} {\bf #1}}

\def\mn#1{{\it Mon. Not. R. astr. Soc.\/} {\bf #1}}

\def\nat#1{{\it Nature\/} {\bf #1}}

\def\ang{\thinspace{\rm \AA}}
\def\approxlt{\lower.2em\hbox{$\buildrel < \over \sim$}}

\def\cm2{{\rm cm}^{-2}}

\def\ha{\ifmmode {{\rm H}\alpha}
        \else {H$\alpha$}\fi}
\def\hno{\ifmmode H_0
    \else $H_0$\fi}

\def\kms{~{\rm km\ s}^{-1}}
\def\la{\ifmmode {{\rm Ly}\alpha}
        \else {Ly$\alpha$}\fi}

\def\omegab{{\Omega_{\rm B}}}
\def\p.{^{\prime\prime}\kern-2.1mm .\kern+.6mm}
\def\pone{^{\prime}\kern-1.05mm .\kern+.3mm}
\def\qnought{\ifmmode q_0
    \else $q_0$\fi}
\def\secpoint{^{\prime\prime}\kern-2.1mm .\kern+.6mm}
\def\sqr#1#2{{\vcenter{\hrule height .#2pt
        \hbox{\vrule width .#2pt height#1pt \kern#1pt
                \vrule width .#2pt}
        \hrule height.#2pt}}}

\def\ten#1{\ifmmode 10^{#1}
    \else $10^{#1}$\fi}

\def\ten#1{\times 10^{#1}}

\def\carbon4{{\ion{C}{4}}}
\def\carbon2{{\ion{C}{2}}}
\def\si4{{\ion{Si}{4}}}
\def\h1{{\ion{H}{1}}}

\def\hi{{\rm H}\thinspace{\sc{i}}}
\def\di{{\rm D}\thinspace{\sc{i}}}

\begin{document}

\title{A high deuterium abundance in the early Universe}
\author{Antoinette Songaila,$^{\ast}$ E. Joseph Wampler$^{\dag}$ \& Lennox L. 
Cowie$^{\ast}$} 

\bigskip\noindent

\leftline{$^{\ast}$\ Institute for Astronomy, University of Hawaii, 2680
Woodlawn Drive, Honolulu, HI 96822}
\leftline{$^{\dag}$\ National Astronomical Observatory, Osawa, Mitaka, Tokyo
181, Japan}

\vskip 1in

\centerline{Accepted for publication in {\it Nature} }

\newpage

{\bf The amount of deuterium relative to hydrogen (D/H) in clouds with close
to primordial abundance seen at high redshift in the spectra of distant
quasars currently provides the best estimate of the baryonic density of the
Universe ($\omegab$).  The first measurements have yielded discrepant values
of D/H both high\cite{schr,carswell,wampler1,rugers1, rugers2} ($\sim
2\ten{-4}$) and an order of magnitude lower.\cite{tytler1,burles} The low
values of D/H imply a high $\omegab$\ that is difficult to reconcile with
determinations of the primordial abundances of other light elements, notably
$^4$He, and with the known number of light
neutrinos.\cite{fields,steigman,hata} We report an independent measurement of
the neutral hydrogen (H\,I) column density in the cloud toward Q1937$-$1009
where one of the low D/H values was obtained.\cite{tytler1} Our measurement
excludes the reported\cite{tytler1} value and we give a lower limit of $D/H >
4\ten{-5}$\ in this system, which implies $\omegab < 0.016$\ for a Hubble
constant of $100~{\rm km\ s^{-1}\ Mpc^{-1}}$.  This new upper limit on
$\omegab$\ relieves the conflict with standard Big Bang nucleosynthesis.}

The accuracy of the measurements of D/H in high redshift quasar absorption
line clouds is heavily
dependent on the neutral hydrogen column density of the absorbing cloud and
different problems arise in different column density regimes.  At sufficiently
low column density,\cite{schr,carswell,wampler1,rugers1, rugers2} absorption
lines of hydrogen high in the Lyman series become unsaturated and permit an
accurate determination of H, but the corresponding weakness of deuterium
increases the likelihood of significant contamination by a chance coincidence
of weak foreground \hi\ absorption.  Alternatively, choosing a higher
column density absorber minimises such deuterium contamination but the
saturation of the Lyman series increases the chance of wrongly estimating the
\hi\ column density.  The two low measurements of D/H
($\sim2.3\ten{-5}$) toward Q1937$-$1009 (ref.~\citen{tytler1}) and
Q1009+2956 (ref.~\citen{burles}) belong in this category.

In the latter case the total amount of neutral hydrogen is hard to estimate
because even the lines high in the Lyman series are nearly opaque.  For the
same reason, the exact velocity distribution of \hi\ and \di\ 
absorption cannot be determined directly but must be plausibly modelled from
the distribution of the weaker absorption of other ions and this procedure can
be problematical.\cite{wampler2} Toward Q1937$-$1009, Tytler et
al.\cite{tytler1} made the minimal assumption that the simplest two-component
model that accounted for low ionization absorption was sufficient to account
for \hi\  and \di\ absorption.  Recent investigation\cite{wampler2}
has shown that alternative models of the absorption in this cloud that have as
much as three times lower \hi\ column density can be found, which
would increase D/H to $7\ten{-5}$, well outside the published\cite{tytler1}
errors.

Regardless of its actual velocity distribution, the total \hi\ column density
associated with a cloud can be found by observing the optical depth of the
corresponding redshifted Lyman continuum.  The different cloud models
proposed\cite{tytler1,wampler2} to account for the absorption toward
Q1937$-$1009 make different predictions for the amount of residual flux below
the Lyman break and these are difficult to distinguish at the required
precision with the existing\cite{tytler1} high resolution Keck HIRES echelle
spectra.  In the violet spectral region, the orders of this spectrograph are
closely spaced and the necessarily short slits limit the amount of sky
available for sky subtraction. We have addressed this difficulty by using
long-slit, lower resolution spectroscopy to measure precisely the residual
flux below the Lyman continuum break.

The measurements were obtained in 1996 August with the Low Resolution Imaging
Spectrograph (LRIS) on the Keck1 10m Telescope on Mauna Kea, Hawaii and
consist of a total of 45 minutes exposure at lower resolution (resolving
power, $R \sim 300$) through a very wide slit ($1\secpoint{5}$) and a higher
resolution $(R \sim 1500$) exposure (40 minutes total) taken through a
$0\secpoint{7}$\ slit.  (Seeing was less than $0\secpoint{8}$\ for both
observations.)  The spectra were flux calibrated with observations of a white
dwarf star taken immediately after the target observations and at nearly
identical zenith angle and declination, and wavelength calibrated with
observations of a Krypton-Mercury lamp in the same configuration.  The
two-dimensional spectral image of the quasar (Figure~1) clearly shows a
residual flux below the continuum break.

The flux-calibrated and sky-subtracted low-resolution spectrum is shown in
Figure~2$a$. 
Both the quasar flux level and the level of the unabsorbed continuum are
needed for calculating the optical depth of the Lyman continuum. For
Q1937$-$1009 the quasar's emission redshift is $z=3.805$, only slightly higher
that the redshift ($z=3.572$) of the absorbing cloud used for measuring the
D/H ratio. The quasar continuum at the $z=3.572$ Lyman limit is therefore in
transition, with the fluctuating absorption from numerous blended lines of
individual Lyman forest clouds becoming increasingly overlayed by the
overlapping Lyman continua of the forest clouds. We have considered two ways
of estimating the continuum level in the $z=3.572$ Lyman continuum region.  The
first was a simple linear extrapolation of the estimated continuum at longer
wavelengths which is shown as the dotted line in Figure~2$a$.  
However, as the figure shows, 
even this very conservative procedure rules out the value of
$N({\rm H\,I})$\ reported in ref.~\citen{tytler1}.  
Tytler reports (private communication) that in the spectral interval $4125\ang
< \lambda < 4175\ang$\ the SNR weighted residual flux $= 0.015 \pm
0.014$. This is not in disagreement with our spectra, as Figure~ 2 shows that
this region suffers particularly heavy absorption from the \la\ forest
lines. However, the total column density adopted in ref. \citen{tytler1}
($N({\rm H\,I}) = 8.7\ten{17}\cm2$) would predict that the residual flux below the
Lyman break = $0.004^{+0.003}_{-0.002}$ which is shown as the dashed line and
error bar, and lies well below the observed spectrum.  In the wavelength
region from $890\ang$\ to $900\ang$\ we measure an average flux of $5.4 \pm
0.4$\ counts compared to an extrapolated average continuum level of 224
counts, where the $1~\sigma$\ error is made by extracting sky regions in the
same way as the quasar.  This corresponds to $\tau = 3.72 \pm 0.06$\ or
$N({\rm H\,I}) = (5.9 \pm 0.1) \ten{17}\cm2$. 

However, the statistical errors are small compared with the systematic error
in the continuum placement and the simple procedure used above overestimates
the value of the optical depth and hydrogen column density.  To provide an
improved estimate, we have modelled the expected continuum level using
statistical Ly$\alpha$ forest data\cite{press,madau}. Such models provide a
good representation of the average forest absorption and we see that they also
match the Q1937$-$1009 forest absorption rather well (dash-dot line in
Figure~2$a$). So long as there are no unrecognized strong absorption systems
with redshifts slightly lower than $z=3.572$, and the model gives an
acceptable representation of the Ly$\alpha$ forest region, the Lyman continuum
absorption required by the line absorption should predict the additional Lyman
continuum absorption decrement to be applied to the quasar spectrum.  In high
redshift Lyman\,$\alpha$ forests, such as the one in Q\,1937$-$1009, the weak
hydrogen clouds are so numerous that they blend into a quasi-smooth
continuum. But Madau et al.\cite{madau} noted that the contribution to
$\tau_{\rm Ly\,\alpha}$ by optically thin absorbers with $10^{16}\leq N({\rm
H\,I})\leq 10^{17}\cm2$\ is highly uncertain. In our calculation of $\tau_{\rm
Ly\,\alpha}$, we retained only the first term in their eq.~5 and reduced its
coefficient by 20\%.  This procedure assumes that there are no high
column density hydrogen clouds in Q\,1937$-$1009 with redshifts near that of
the $z=3.572$ system. This assumption is justified both by the good match of
the statistical model to Lyman\,$\alpha$ forest region of our spectra and by
the finding of Tytler \& Burles\cite{tytler2} that the column densities of the
three highest column density hydrogen clouds with redshifts to the blue of the
D/H absorption system have $15.06\leq \log\,N_{\rm H}
\leq15.7$. As the combined absorption produced by these three systems is 
$\leq 10$\% of Tytler's predicted absorption for the much stronger $z=3.572$
D/H system it can be neglected in comparison to the uncertainties inherent in
setting the appropriate level for the un-absorbed quasar continuum, and the
accurate determination, with only low resolution spectra, of the $z=3.572$
Lyman continuum absorption in the presence of strong line absorption from
other redshift systems.  This fit reduces the extrapolated continuum level at
$890 - 900\ang$\ to 126 counts, which gives an optical depth of $3.15 \pm
0.06$\ or $N({\rm H\,I}) = (5.0 \pm 0.1) \ten{17}\cm2$.

Figure~2$b$
shows a linear plot of the moderate resolution spectrum in the region near the
$z=3.572$ Lyman limit. Again the two continuum estimates are shown. Here
Ly$\alpha$ forest lines can be seen cutting into the residual flux both above
and below the break. The residual flux levels from the moderate and low
resolution spectra are in satisfactory agreement when the moderate resolution
data is smoothed to the low resolution.  Here we obtain $\tau_{\rm cont}=3.1$
(linear extrapolation of the quasar continuum) or $\tau_{\rm cont}=2.4$
(Ly$\alpha$ forest model continuum) where we have made the measurement at the
rest-frame wavelength of $890\ang$\ in the high resolution spectrum.  These
optical depths translate to $3.8\ten{17}\cm2<N({\rm H~I})<4.9\ten{17}\cm2$ for
the total hydrogen column density in the $z=3.572$ absorption system.  

Based on the above analysis of the two spectra, we consider a value of
$5\ten{17}\cm2$\ to be a reasonable maximum neutral hydrogen column density
that can be associated with the measured\cite{tytler1} deuterium column
density of $N({\rm D~I}) = 2\ten{13}\cm2$. This then gives a formal minimum
for D/H of $4\ten{-5}$.  


However, this new determination of total \hi\ column density cannot
distinguish different \hi\ or \di\ velocity components and so cannot measure
the D/H ratio in any one component. Spectra with a resolution R$\geq6\ten{4}$
are needed to resolve the velocity structure of the low-ionization metal line
clouds in order to settle the issue of the number and relative strengths of
the component clouds within the $z=3.572$ system. More precise cloud
information might show that a considerable fraction of the accompanying
hydrogen could be located in the red wing of the line.  If so, the
corresponding deuterium would be obscured by overlying hydrogen, and D/H
ratios even as high as $2\ten{-4}$\ could then be accommodated by the data.
We would like to emphasise that the highly uncertain and model-dependent
nature of the systematic errors involved in these measurements of deuterium
makes the assignment of formal systematic errors very difficult. Since this is
likely to remain the case as new measurements accumulate, we have
chosen rather to emphasise a reliable lower bound, while giving some idea
through modelling of the possible uncertainty.  

We would therefore recommend adopting a range of $4\ten{-5} < D/H <
2.4\ten{-4}$\ as the most conservative current interpretation of the results
of determining D/H in the high redshift quasar absorption line
systems.  The density parameter, $\Omega_{\rm B}$, is then constrained to be
$0.005 <\Omega_{\rm B}\,h^2 < 0.016$, where the Hubble constant, $\hno
=100\,h\kms~{\rm Mpc}^{-1}$.  Unlike the lower D/H value of $2.3\ten{-5}$,
this range is broadly consistent with measurements of $^7$Li\ and $^4$He
[ref.~\citen{fields}] and eliminates much of the evidence that
suggests\cite{steigman}  that
there might be a crisis in the Standard Big Bang Nucleosynthesis model.  It
remains true that the upper end of the range is in better agreement with
$^4$He and $^7$Li measurements whereas the lower value is easier to reconcile
with models of Galactic chemical evolution tied to the locally measured value
of D/H, as well as with local estimates of $\omegab$.  This issue will be
resolved only when further measurements of D/H\ narrow the range.

\smallskip

We are grateful to E. Hu for her assistance in obtaining the data.  The
authors were visiting astronomers at the W.  M. Keck Observatory, jointly
operated by the California Institute of Technology and the University of
California.  The research was supported at the University of Hawaii by the
State of Hawaii.

\newpage

\begin{figure}[H]
\vbox to3.0in{\rule{0pt}{3.0in}}
\includegraphics{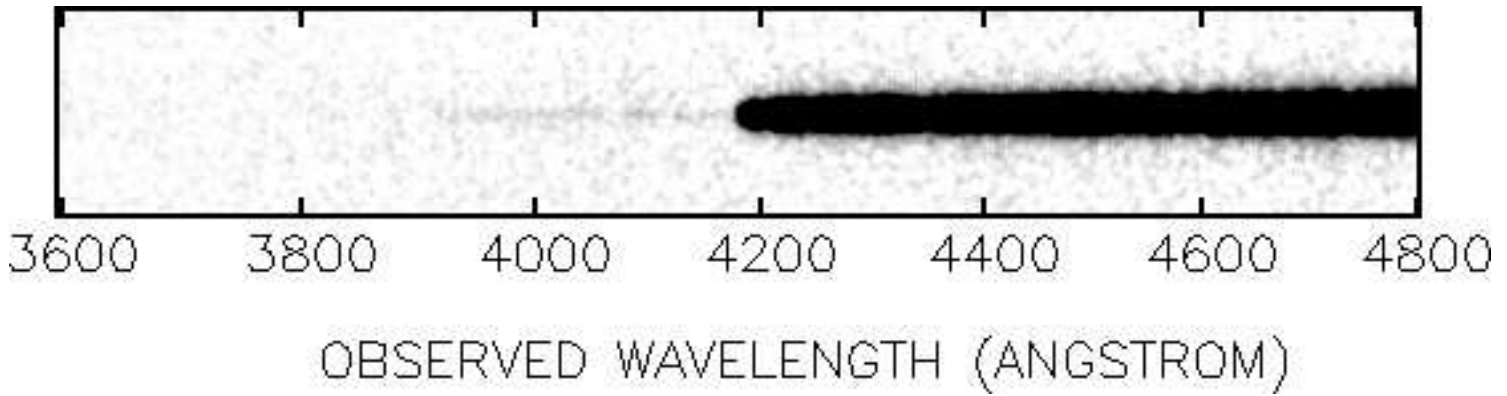}
\caption{The two-dimensional spectrum of Q1937$-$1009 is shown in a grey-scale 
representation with the dispersion direction along the horizontal, covering
the wavelength range from $3600~\rm\AA$\ to $4800~\rm\AA$, and the spatial
direction along the vertical ({$\pm 8\secpoint{5}$} about the quasar).  The
Lyman continuum break is positioned at the center of the picture with the
residual flux at shorter wavelengths running to the left.  The spectrum was
formed from 3 15-minute exposures obtained with the $300\ell~{\rm mm}^{-1}$\
grating on the LRIS spectrograph on the Keck1 10m telescope using a
\protect{$1\secpoint{5}\times 172\arcsec$} slit.  The resulting spectrum had 
$R = 300$.  The quasar was moved to a different position along the slit
between each exposure.  The frames were sky-subtracted using the median of
positionsrepresentation with the dispersion direction along the horizontal, covering
the wavelength range from $3600~\rm\AA$\ to $4800~\rm\AA$, and the spatial from \protect{$8\secpoint{5}$} to
\protect{$4\secpoint{5}$} from the quasar on either side of the quasar.  The
three frames were then median added to eliminate cosmic rays, and finally
calibrated to $F_{\nu}$\ using observations of the white dwarf photometric
standard star Feige 110, obtained in the same configuration.  The spectrum was
wavelength calibrated using a third-order polynomial fit to observations
of a Krypton-Mercury lamp, and the calibration checked against the Balmer line
positions in the white dwarf.}
\end{figure}

\begin{figure}[H]
\caption{(Following page) (a)---The spectrum (flux per unit frequency in
arbitrary units) of Q1937$-$1009 extracted from the two-dimensional spectral
image of Figure~1 is shown as a function of the rest wavelength in the $z =
3.572$\ quasar absorption line frame, showing the position of the Lyman
continuum break.  The solid line shows a fit to the upper envelope of the
spectrum redward of the $\la$\ emission line, excluding known emission lines.
The dash--dot line shows the level of the corresponding continuum expected
blueward of the $\la$\ emission line from ref.~\citen{madau} .  The dotted
line shows a simple linear fit to the upper envelope of the continuum just
above the Lyman continuum break. The dashed line shows the expected residual
flux from the prediction of ref.~\citen{tytler1}.  The error bar shows the
uncertainty in that prediction, including both the quoted statistical and
systematic errors.  Both this spectrum and the higher resolution
spectrum (below) were oversampled and are plotted without smoothing. \quad
(b)--- The higher resolution spectrum ($R \sim 1500$) is shown as flux per
unit frequency as a function of rest wavelength in the quasar absorption line
frame.  The spectrum was extracted in the same way as the lower resolution
spectrum was, and its spectrophotometry found to be in good agreement with the
lower resolution wide-slit spectrum.  This higher resolution shows the rich
forest of Lyman alpha lines.  We show again both the linear (dotted) and
ref.~\citen{madau} (dash--dot) continuum fits.  We have measured the optical
depth at $890\ang$\ in the rest frame, where the signal-to-noise is still high
in the residual spectrum, and find an optical depth, $\tau = 2.4$\ for the
continuum of ref.~\citen{madau} and $\tau = 3.1$\ for the linear fit.}
\end{figure}

\begin{figure}[H]
\vbox to9.0in{\rule{0pt}{9.0in}}
\includegraphics{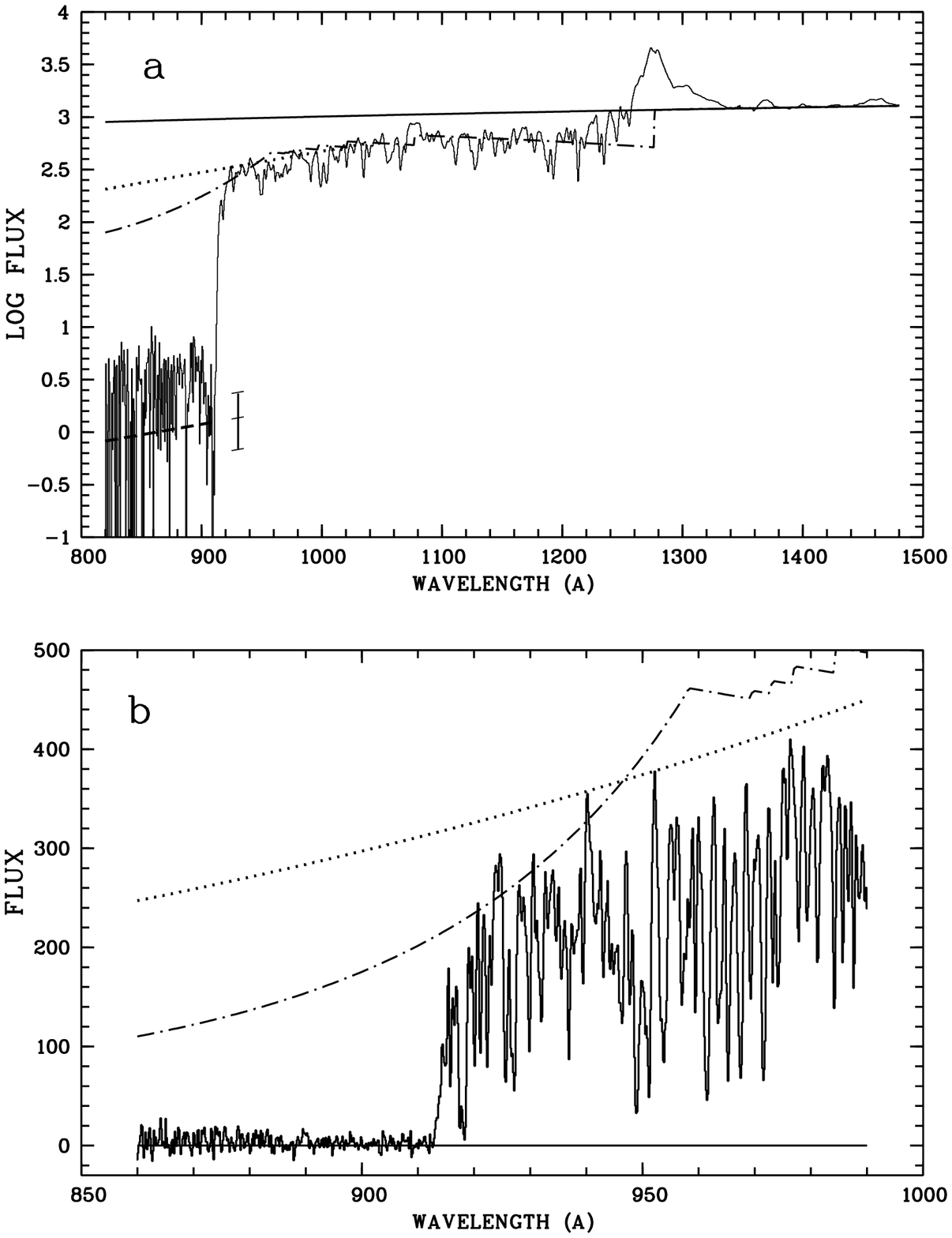}
\end{figure}


\begin{thebibliography}{50}

\bibitem {schr} Songaila, A., Cowie, L. L., Hogan, C. J. \& Rugers, M.\
\nat{368}, 599--603 (1994).
\bibitem {carswell} Carswell, R. F. et al.\ \mn{268}, L1--L4 (1994).
\bibitem {wampler1} Wampler, E. J. et al.\ \aa{}, in the press (1996).
\bibitem {rugers1} Rugers, M. \& Hogan, C. J.\ \apj{459}, L1--L4 (1996).
\bibitem {rugers2} Rugers, M. \& Hogan, C. J.\ preprint (1996).
\bibitem {tytler1} Tytler, D., Fan, X.-M. \& Burles, S.\ \nat{381}, 207--209
(1996). 
\bibitem {burles} Burles, S. \& Tytler, D.\ {\it Science}, submitted (1996).
\bibitem {fields} Fields, B. D., Kainulainen, K., Olive, K. A. \& Thomas, D.   
{\it New Astronomy}, {\bf 1(1)}, 77--96 (1996).
\bibitem {steigman} Steigman, G. {\it The Crisis Confronting Standard Big Bang
Nucleosynthesis}, to appear in {\it Critical Dialogs in Cosmology} (Princeton
University Press) (1996).
\bibitem {hata} Hata, N., Steigman, G., Bludman, S. \& Langacker, P.\ {Phys.
Rev. D}, in the press (1996).
\bibitem {wampler2} Wampler, E. J.\ \nat{383}, 308 (1996).
\bibitem {press} Press, W.H. \& Rybicki, G. \apj{414}, 64--81 (1993).
\bibitem {madau} Madau, P. {\em et al.} Preprint (Astro-Ph/9607172).
\bibitem {tytler2} Tytler, D. \& Burles, S.  in {\em Origin of Matter
and Evolution of Galaxies in the Universe '96}, ed. T. Kajino, Y. Yoshii, 
S. Kubono, (World Scientific: Singapore), in press (1996) (Astro-Ph/9606110).

\end{thebibliography}
\end{document}